# Suppression of photo-oxidation of organic chromophores by strong coupling to plasmonic nanoantennas

Battulga Munkhbat[1], Martin Wersäll[1], Denis G. Baranov[1], Tomasz J. Antosiewicz[1,2], and Timur Shegai[1*]

[1]Department of Physics, Chalmers University of Technology, 412 96, Göteborg, Sweden

[2]Centre of New Technologies, University of Warsaw, Banacha 2c, 02-097 Warsaw, Poland

e-mail: timurs@chalmers.se

KEYWORDS: Strong coupling; photochemistry; organic dyes; plasmonics; single nanoparticle spectroscopy; J-aggregates;

**ABSTRACT**

Intermixed light-matter quasiparticles – polaritons – possess unique optical properties owned to their compositional nature. These intriguing hybrid states have been extensively studied over the past decades in a wide range of realizations aiming at both basic science and emerging applications. However, recently it has been demonstrated that not only optical, but also material-related properties, such as chemical reactivity and charge transport, may be significantly altered in the strong coupling regime of light-matter interactions. Here, we show that a nanoscale system, comprised of a plasmonic nanoprism strongly coupled to excitons in J-aggregated form of organic chromophores, experiences modified excited state dynamics and therefore modified photo-chemical reactivity. Our experimental results reveal that photobleaching, one of the most fundamental photochemical reactions, can be effectively controlled and suppressed by the degree of plasmon-exciton coupling and detuning. In particular, we observe a 100-fold stabilization of organic dyes for the red-detuned nanoparticles. Our findings contribute to understanding of photochemical properties in the strong coupling regime and may find important implications for the performance and improved stability of optical devices incorporating organic dyes.

# INTRODUCTION

Strong light-matter interactions in nanoscale objects have attracted considerable research attention recently. These interactions manifest themselves in emergence of new hybridized eigenstates of a strongly coupled system evidenced by coherent Rabi oscillations between the matter and the photonic subsystems (*1, 2*). This behavior is strikingly different from weak light-matter coupling, which results in the conventional spontaneous emission process accelerated by the cavity through the Purcell effect. Strong coupling between optical and plasmonic cavity resonances and matter excitations have been recently realized in systems involving molecular (*3-14*) and semiconductor (*15, 16*) excitons, as well as vibrational transitions (*17-19*). Remarkably, strongly coupled systems have been demonstrated to enable ultrafast optical switching and single-photon nonlinearities (*20-22*).

It is important to point out that in the strong coupling regime the photonic and excitonic components of the system cannot be treated as separate entities, as they form new polaritonic eigenstates possessing both light and matter characteristics. For this reason, not only electromagnetic, but also material (microscopic) properties of a strongly coupled system can be modified. In particular, it has been shown recently that chemical reactivity (*19, 23-25*) and charge and exciton transport (*26-28*) can be modified under strong coupling conditions. Strong coupling has been demonstrated to tune the work-function (*29*), as well as to control the Stokes shifts of hybridized molecules (*30*). Feist *et al.* (*24*) have theoretically described an increased stability in the strong coupling regime for a photo-isomerization reaction. Herrera *et al.* (*25*) showed that strong coupling can modify molecular reaction pathways via polaron decoupling. These observations open up new possibilities to employ strong coupling for controlling chemical reactivity and other material-related properties. However, only a small number of chemical reactions in the strong coupling regime have been experimentally studied to date.

Here, we demonstrate that strong coupling between organic molecules and plasmonic nanocavities can significantly alter one of the most important classes of photo-chemical processes – the photobleaching reaction. Photobleaching limits applications of organic electronics devices such as, organic solar cells, electroluminescent devices, organic dye lasers and molecular fluorescence, as it leads to irreversible photodegradation of organic molecules (*31, 32*). Therefore, it is important to seek for means to control and prevent photobleaching. Here, we observe that upon coupling of J-aggregates to a plasmonic nanocavity, the photobleaching rate can be dramatically suppressed (up to 100-fold), what is evidenced by recording the evolution of the scattering response of the coupled systems over time. Moreover, our study reveals complex relations between the photobleaching rate and the plasmon-exciton detuning as well as the excitation conditions. These results point to a beneficial impact of strong coupling on photochemical properties and pave the way towards molecular optical nanostructures with increased photostability.

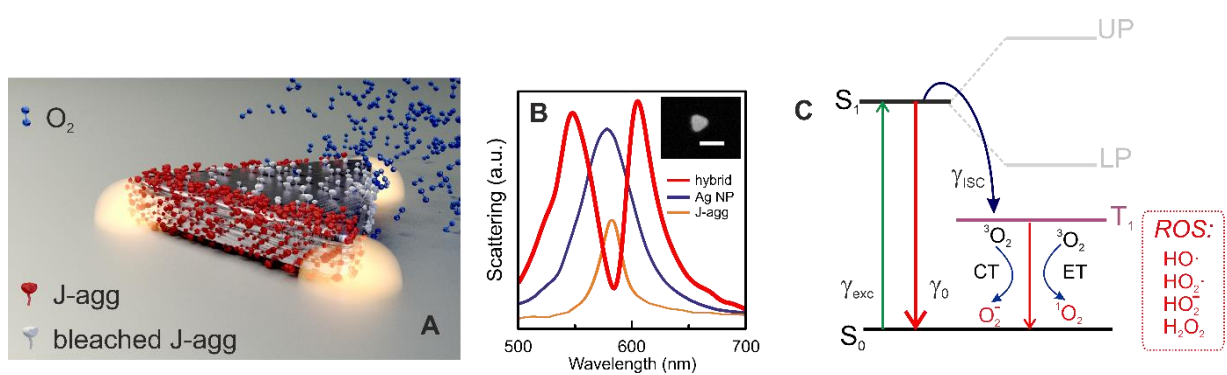

**Fig. 1. Sketch and schematic diagram of photobleaching reaction in a strong-coupling system.** (A) Graphic sketch of the nanoscale system. (B) Dark-field scattering spectra of the strongly coupled hybrid system (red), as well as scattering spectrum from uncoupled J-aggregates (orange) and uncoupled individual plasmonic nanoprism (blue). The inset shows an SEM image of the corresponding nanoprism (scale bar 100 nm). (C) Schematic diagram of a photobleaching reaction in the uncoupled molecular system and its possible modification in the strong coupling regime (light grey lines indicate upper and lower polaritonic states).

**RESULTS**

*Concept.* It is generally accepted that the mechanism of photobleaching involves photodynamic interactions between the excited triplet state of dye molecules and atmospheric triplet oxygen ($^3O_2$). For high quantum yield intersystem crossing (ISC) fluorophores, a significant population transfer from the excited singlet state ($S_1$) to the long-lived triplet state ($T_1$) can occur (see Fig. 1C). Because the lifetime of the triplet state is long, the excited molecules have a higher chance to interact with environmental oxygen and thus trigger photobleaching. The interaction between molecular oxygen and triplet excitons occurs via energy or charge transfer mechanisms and leads to generation of several reactive oxygen species (ROS), such as singlet oxygen ($^1O_2$) and hydrogen peroxide ($H_2O_2$), which are highly unstable and therefore can chemically damage fluorophores and their surrounding (*33-35*). When such a system is brought into interaction with the resonant cavity (see Fig. 1A), the relaxation pathways and therefore photobleaching may be drastically affected. Such interaction can be weak (i.e. Purcell regime) or strong (i.e. vacuum Rabi splitting). In both cases, the photochemical reaction can be modified, but when the system enters the strong coupling regime, the relaxation pathways of the hybrid can be modified to a very large degree. This is because strong coupling, as opposed to the weak counterpart, is a coherent effect that manifests itself through a buildup of collective states encompassing all participating molecules.

Considerable progress on the suppression of photobleaching has been achieved recently using inert ambient as well as oxygen scavenger reagents (*36, 37*). Furthermore, plasmonic nanoantennas were shown to improve the photostability of organic dyes by quenching long-lived triplet states (*38-40*). Other important examples are known from surface-enhanced Raman and enhanced fluorescence experiments, where stability of single molecule trajectories can often exceed several thousands of seconds, while the same molecules in free space typically photobleach in a matter of seconds. It is believed that enhanced spontaneous emission in the form of efficient competition with ISC is responsible for these observations (*41-43*). However, these experiments were performed in the weak coupling regime, whereas the potential of strong coupling for the purpose of photostability has not been explored.

In this study we suppress the photobleaching reaction by depleting the triplet state population in the strong coupling regime. Because polaritonic states are coherent mixtures of plasmons and excitons, their corresponding lifetime is extremely short (~10 fs) due to partial plasmonic character. The typical ISC rate is many orders of magnitude slower ($10^{-8} - 10^{-3}$ s) than that (*44, 45*), and therefore the polaritonic states are not able to populate the triplet efficiently. Thus involvement of polaritonic states must be highly useful for suppressing the photobleaching reaction. The goal of this study is to elucidate the quantitative aspects of this problem.

***System Under Study.*** Our study is focused on the suppression of photobleaching of strongly coupled organic J-aggregate molecules to a plasmonic nanoantenna. The hybrid system is formed by a silver nanoprism surrounded by J-aggregates of 5,5',6,6'-tetrachloro-di-(4-sulfobutyl) benzimidazolocarbocyanine (TDBC) organic dye molecules (see Fig. 1A and Methods for further details). The optical response of the individual hybrid system, uncoupled plasmonic nanoprism and uncoupled J-aggregates are furthermore characterized by their scattering spectra in Fig. 1B, together with a scanning electron microscopy (SEM) image of the corresponding individual silver nanoprism. The hybrid system shows Rabi splitting of the order of ~200 meV which exceeds not only polariton, but also uncoupled plasmon resonance width. We have investigated several tens of individual hybrid systems, which show reproducible Rabi splitting exceeding the plasmon linewidth. We thus undoubtedly conclude that the system under study is in the strong coupling regime, in agreement with our recent observations (*13, 14*). We also note that the hybrid polaritonic states in our system are formed by mixing plasmonic and excited singlet states of J-aggregates. At the same time, the triplet state of J-aggregates has an extremely weak oscillator strength and so strong coupling between plasmonic field and the triplet is not feasible.

***Effect of Rabi Splitting on photobleaching.*** To investigate the crucial impact of Rabi splitting on the suppressed photobleaching of the hybrid system, the time-resolved degradation of several coupled systems under incident light/photoexcitation was studied (see Fig. 2). To visualize photobleaching rates of the system, dark-field (DF) scattering spectra of individual hybrid systems as a function of exposure time were recorded upon irradiation with a spectrally flat broadband light beam at a constant intensity (see Methods). Figs. 2A-E show the bleaching time-resolved DF scattering spectra as false color plots for five individual hybrid structures, with similar plasmon-exciton detuning but different initial Rabi splitting values of 209, 200, 188, 170, and 154 meV, respectively. Importantly, electron microscopy data shows that all of the hybrid systems are composed of an individual silver nanoprism and J-aggregated molecule around (see insets in Figs. 2A-E). Moreover, the size of the particles and their plasmon-exciton detuning is very similar, implying that the absorption cross-sections are approximately the same. More examples are shown in fig. S1 in the section S1. The DF scattering spectra of the hybrid systems at 0, 240 and 480 s of bleaching time are displayed as blue, dark-brown, and black-dashed line curve in Figs. 2A-E, respectively. For comparison, Fig. 2F also shows the evolution of scattering from uncoupled J-aggregates under the same experimental conditions. Rabi splitting decreases with time for all samples indicating degradation of organic molecules under optical excitation.

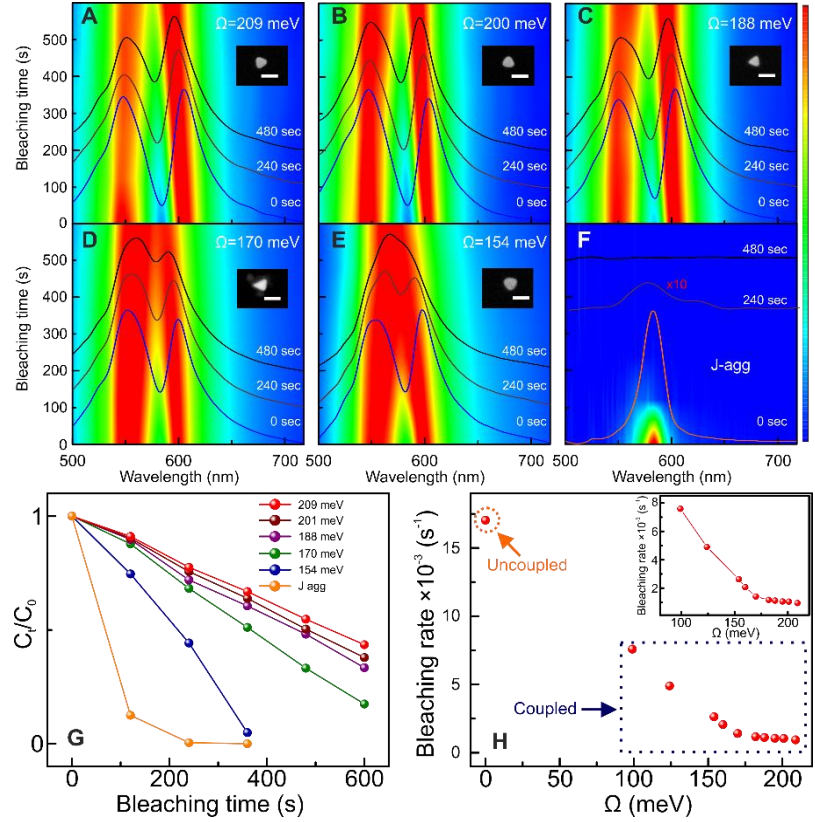

**Fig. 2. Photobleaching as a function of Rabi splitting.** (A-E) Evolution of the dark-field scattering spectra of strongly coupled hybrid systems with different initial Rabi splitting energies of 209, 200, 188, 170, and 154 meV, respectively. Insets show the corresponding SEM images of the plasmonic silver nanoprisms (scale bar 100 nm). (F) Uncoupled J-aggregates as a function of irradiation/bleaching time under broad-band excitation. (G) Photobleaching kinetics for the strongly coupled hybrid systems with different Rabi-splitting energies of 209 meV (red), 200 meV (brown), 188 meV (violet), 170 meV (green), and 154 meV (blue), compared to the uncoupled bare J aggregates (orange). (H) Photobleaching rates for hybrid systems vs uncoupled J-aggregates as a function of Rabi-splitting energy. The inset shows a zoomed-in plot of the photobleaching rates for the coupled systems.

The data presented in Figs. 2A-E allows extracting the relative change in the concentration of active organic molecules $C(t)/C(0)$ over time. In order to do that, we recall that the value of Rabi splitting between is given by $\Omega_R = 2\sqrt{(\omega_+ - \omega_0)(\omega_0 - \omega_-)}$ with $\omega_\pm$ being the frequencies of the upper (UP) and lower polariton (LP) that can be inferred directly from the recorded DF spectra and $\omega_0$ is the exciton resonance energy. Then the relative change in the concentration of active molecules is estimated as $C(t)/C(0) \approx [\Omega_R(t)/\Omega_R(0)]^2$, since $\Omega_R \sim \sqrt{N/V} = \sqrt{C}$ where $N$ is the number of active molecules residing within the mode volume $V$.

The obtained curves clearly show one of the central results of our work: photobleaching of the hybrid systems *occurs at a much slower rate* than the uncoupled systems (see Fig. 2G). Moreover, the hybrid system becomes more stable against photobleaching as the Rabi splitting is increased. Even after 600 s of continuous irradiation, the hybrid systems with the three largest Rabi splittings still show two hybridized modes in the DF scattering, indicating that many molecules still actively participate in the coupling process. In contrast to that, uncoupled J-aggregates are rapidly bleached showing ~88% decrease in the intensity of DF scattering spectra within the first 120 s. Importantly photobleaching dynamics is almost linear in the hybrid cases, while it is exponential in the uncoupled J-aggregate case on the time intervals studied. This observation alone implies that photobleaching of hybrid systems is considerably slowed down. We will come back to this point later, when discussing the kinetics.

Observations in Figs. 2A-F show that strong coupling indeed influences the photo-chemical processes, likely by alternating triplet state population and relaxations pathways. We stress that the particle sizes used were similar and thus their extinction cross-sections and mode volumes are comparable (see SEM insets in Fig. 2). The experiments were performed under exactly the same illumination conditions. Therefore, these results cannot be explained by variation of particle's extinction and/or experimental conditions.

For a further quantitative comparison of the hybrid systems, we analyze the photobleaching kinetics curves in Fig. 2G. We extract the bleaching rates as the slope of the kinetic curves assuming quasi-linear dynamics of the active molecules concentration (see Fig. 2H). The extracted bleaching rates for the hybrid samples clearly depend on the Rabi splitting, such that stronger coupled systems are more stable while weaker coupled ones are less stable (see inset in Fig. 2H). The range of Rabi splittings studied in these experiments was ~100-220 meV. By extrapolating the bleaching rate to Rabi splittings in between 0-100 meV, we could expect that the weakly coupled systems would be less stable than even the most unstable strongly coupled ones (see Fig. 2H). To understand this, we rely on several theoretical works where the transition dynamics between polaritonic and dark states was studied (*25, 46-50*). Within the strong coupling picture, the photobleaching rate is determined by a sequence of processes involving excitation of the UP state, its subsequent relaxation to the ground state and to the incoherent states, and decay of the incoherent states into the lower energetic states, which include both LP and triplet states. We come back to this point in more detail, when discussing the photobleaching mechanism.

In concluding this section, we note that Fig. 2 evidently demonstrates suppression of photobleaching rate due to strong coupling. Moreover, for larger Rabi splitting this suppression is enhanced. These experiments have been performed with zero-detuning hybrids and under a broadband optical excitation (350-715 nm). In order to shed more light on this problem, we have performed several additional experiments, where detuning between plasmons and excitons is varied and narrow band optical excitation across the visible range were used.

***Effect of detuning on photobleaching.*** In order to get a deeper insight into the effect of different parameters on the photobleaching of the hybrid systems, we have investigated the influence of plasmon-exciton detuning on photobleaching dynamics. In Fig. 3, we show false color scattering plots for several individual hybrid nanostructures with different plasmon-exciton detunings. DF scattering spectra in color plots are shown using logarithmic scale in order to visualize both polariton peaks. More examples of the hybrid systems with different plasmon-exciton detunings are shown in fig. S2 in the section S2. The examples of low and high blue-detuned hybrid systems are displayed in Fig. 3A,B. For comparison, Fig. 3C,D show examples of low and high red-detuned hybrid systems. We observe that strongly-coupled systems exhibiting red-detuning of the plasmon resonance with respect to the exciton resonance experience much slower photobleaching than the blue detuned ones (see Fig 3E,F).

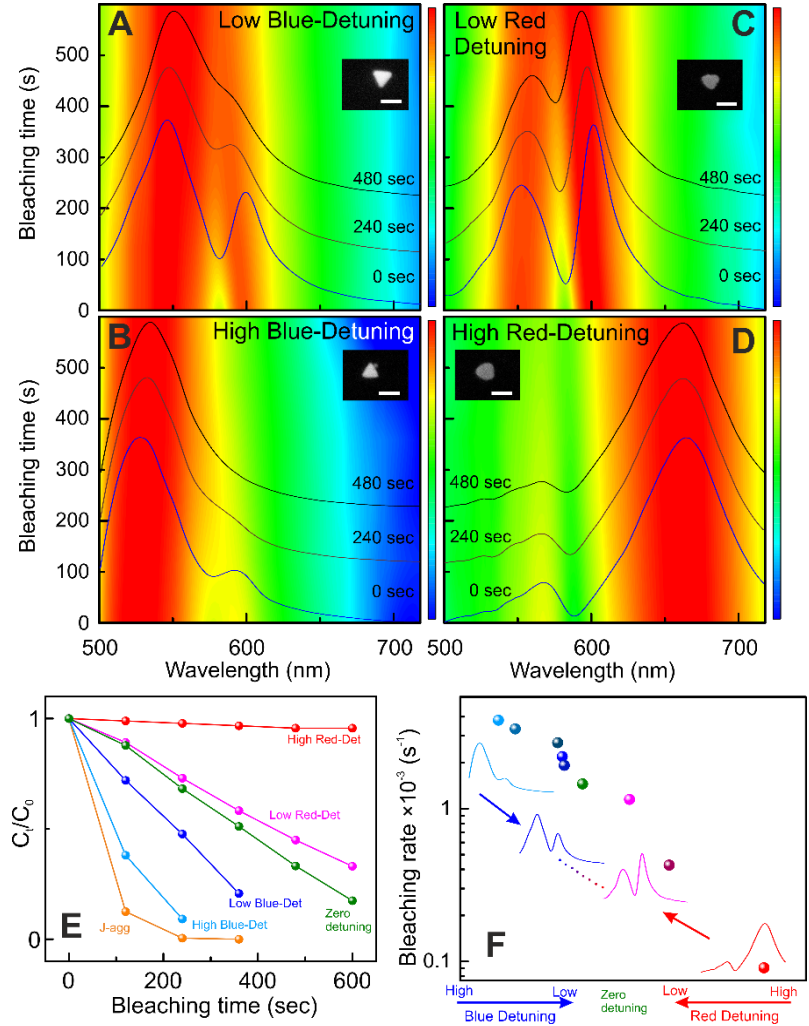

**Fig. 3. Photobleaching as a function of plasmon-exciton detuning.** (A-D) Evolution of the dark-field scattering spectra of strongly coupled hybrid systems with time under broadband excitation for different blue- (A, B) and red-detuned (C, D) plasmon resonances with respect to the exciton resonance. Insets show corresponding SEM images of the individual silver nanoprisms (scale bar is 100 nm). (E) Photobleaching kinetics and (F) bleaching rates for the strongly coupled hybrid systems with different plasmon-exciton detuning. Note the logarithmic scale.

As can be seen in Fig 3A,B, the low blue-detuned case of a hybrid system shows slower photobleaching than the highly blue-detuned hybrid system. The lower polariton peak in the high blue-detuned hybrid system disappears within first 240 s of light irradiation, while the low blue-detuned one still shows splitting. Apparently, the high blue-detuned hybrid system exhibits a much increased photobleaching rate; nonetheless it is still slower than photobleaching of uncoupled molecules (see Fig. 3E,F).

In contrast to the blue-detuning cases, photobleaching occurs much slower in the red-detuned hybrid systems as shown in Fig. 3C,D. In particular, the highly red-detuned hybrid exhibits Rabi splitting even after 480 s of continuous irradiation. As the red-detuning is increased, the photobleaching in the strongly coupled systems is suppressed more significantly, as shown in Fig. 3E,F. In particular, for highly red-detuned Ag nanoprisms, the photobleaching rate is suppressed more than 100-fold in comparison to uncoupled J-aggregates! This is one of the central observations of our study. In the following sections, we focus on the explanation of these observations.

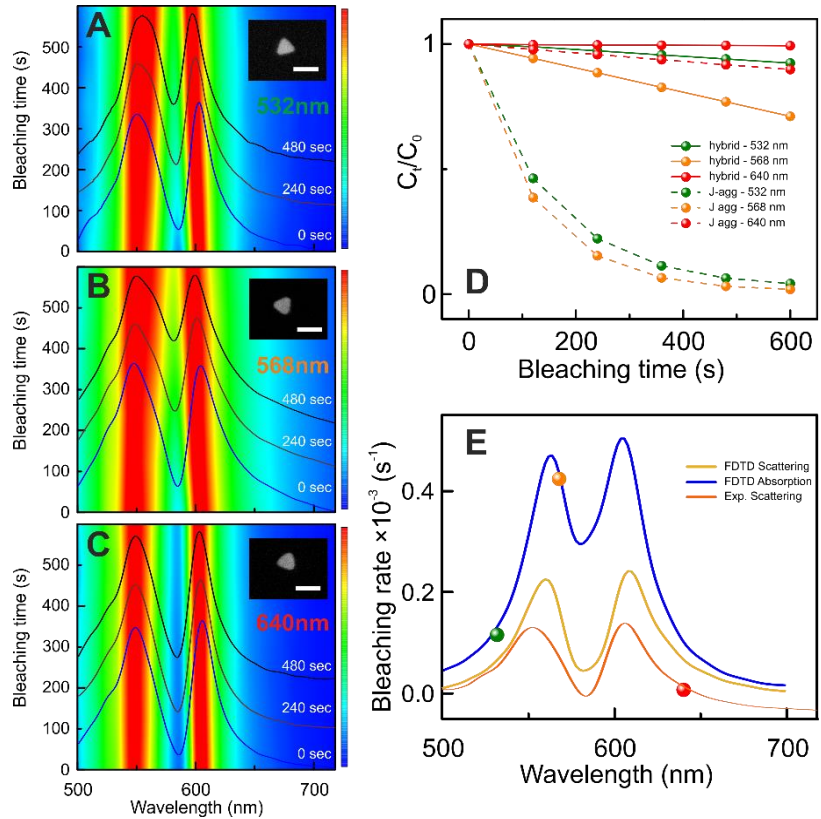

**Fig. 4. Photobleaching as a function of excitation wavelength.** (A-C) Dark-field scattering spectra of three identical hybrid systems (A-C) at three different excitation resonances of 532, 568, and 640 nm, respectively. The insets in (A-C) show corresponding SEM images of the individual silver nanoprisms (scale bar 100 nm). (D) Photobleaching kinetics for the three nearly identical hybrid systems excited using of 532 nm (green solid line), 568 nm (orange solid line), and 640 nm (red solid line) light sources. The corresponding dashed lines show the data for uncoupled J-aggregates. (E) Photobleaching rates of strongly coupled hybrid systems as a function of excitation wavelength for zero-detuned hybrids. The red solid line shows a representative example of DF spectrum for the coupled system. Calculated absorption and scattering spectra are also shown.

***Effect of optical excitation.*** In order to evaluate contributions of each polaritonic states, namely upper and lower polaritons, in the photobleaching we performed experiments using narrow band variable wavelength optical excitations (see Fig. 4). Three different hybrid systems with comparable initial Rabi splitting and plasmon-exciton detuning were chosen and excited using narrow band light sources centered on 532, 568, and 640 nm aiming at selective excitation of the UP, uncoupled molecules and the LP states, respectively (see Methods for further details). The evolution of DF spectra for each system is presented in Figs. 4A-C. The relative change in concentration of active molecules over time is extracted from the DF spectra for all three different excitation wavelengths, as shown in Fig. 4D. In addition, the bleaching kinetics of uncoupled J-aggregates under the same illumination conditions are shown for comparison. Lastly, the bleaching rates for three excitation wavelengths are extracted from Fig. 4D and plotted against the initial DF and calculated absorption spectra of the coupled system (see Fig. 4E).

We observe that pumping the LP with 640 nm excitation causes the slowest bleaching (red-solid line in Fig. 4D), whereas pumping at the wavelength of 568 nm causes the fastest bleaching of the coupled system. The reason for the fastest bleaching in this case is likely a better spectral overlap with the UP state absorption, which in turn populates the triplet state via long-lived intermediates. However, the strongly coupled system, pumped with 568 nm excitation (orange-solid), still exhibits much slower photobleaching compared to uncoupled J-aggregates (orange-dashed) under same experimental condition, as shown in Fig. 4D. When we excite the upper polariton using excitation wavelength of 532 nm (green-solid), the resulting photobleaching rate is slower than at 568 nm, but faster than the direct excitation of lower polariton at 640 nm. This is because selective excitation of the LP cannot contribute directly to photobleaching due to its very short lifetime (~10 fs), while the UP can contribute to photobleaching due to inclusion of incoherent states into its relaxation pathways. The strongly coupled systems exhibit slower photobleaching compared to uncoupled J-aggregates under the same experimental conditions for all studied excitations (see Fig. 4D).

In Fig. 4E, we observe that resonant excitation of the UP is more harmful for the coupled system than non-resonant high energy excitation (likely due to higher extinction cross-section of the former). We thus argue that the only efficient pathway for photobleaching of the coupled system is the resonant excitation of the UP state and its subsequent decay into the incoherent states. With this in mind, we now turn to the in-depth discussion of the photobleaching mechanism of the strongly coupled system.

***Photobleaching mechanism.*** The photobleaching mechanism can formally be described by a set of differential rate equations (see section S4). The relaxation pathways of the coupled system are schematically shown in Fig. 5A. The ground state, upper- and lower polariton energies are indicated as G, UP and LP. The UP and LP states are separated by the Rabi splitting. Moreover, the *incoherent states* (or exciton reservoir) are denoted as D. These states arise as a result of structural and orientation disorder in organic microcavities (*46*). The wavy dashed lines joining the UP to D states and D states to the LP illustrate non-radiative energy transfer processes, which are described by the rates $\gamma_{UP \to D}$ and $\gamma_{D \to LP}$, correspondingly. The full lines from the UP and LP to the G represent the direct decay of polaritonic states - $\gamma_{UP}$ and $\gamma_{LP}$, correspondingly. All these states are singlet in terms of its electron spin. In addition, there is an exciton triplet state, whose lifetime is relatively long. This state is populated via intersystem crossing, $\gamma_{ISC}$, and can decay via a combination of radiative and non-radiative processes, which we denote as $\gamma_T$. It also can decay due to interaction with triplet oxygen from the environment, described by the second order chemical reaction process - $k[\,^3O_2]$. In principle, LP may be found below the triplet state and thus activate the reversed ISC, as well as to enhance triplet relaxation via the Purcell effect. However, in typical organic dyes, the energy gap between singlet and triplet is 0.5-1 eV (*51, 52*), which is significantly above the Rabi splitting in the current study. Therefore these mechanisms are unlikely to affect the dynamics in the present study. We note that the triplet state in J-aggregates is nearly unaffected by dipole-dipole interactions because of the weak oscillator strength of the triplet state. Therefore the J-aggregate triplet appears at the same energy as the triplet of the monomer band – 0.5-1 eV below the monomer singlet state (*53*). We also note that in Fig. 5A, we deliberately ignored the vibronic fine structure of all involved states. This was done because TDBC J-aggregates are well-known to have a small Stokes shift.

Under quasi steady-state operation, the equilibrium population of all intermediates, including the T, D, LP and UP states is reached. Thus, we may write $\gamma_{ISC} n_D = (k[\,^3O_2] + \gamma_T) n_T$, where $n_D$ and $n_T$ denote equilibrium population of incoherent and triplet states, correspondingly. Within this model, the photobleaching process can be described by a gradual degradation of the strongly coupled molecules in accordance with: $\frac{dN}{dt} = -\gamma_{bl} N$, where $N$ is the number of molecules participating in the coupling process and $\gamma_{bl}$ is the photobleaching rate, which in turn reads (detailed derivation is provided in section S4):

$$\gamma_{bl} = \gamma_{exc} \phi_{bl} \qquad \text{Eq. (1),}$$

where $\phi_{bl} = \left(\frac{\gamma_{UP \to D}}{\gamma_{UP} + \gamma_{UP \to D}}\right)\left(\frac{\gamma_{ISC}}{\gamma_{D \to LP} + \gamma_D + \gamma_{ISC}}\right)\left(\frac{k[\,^3O_2]}{k[\,^3O_2] + \gamma_T}\right) \ll 1$ is the quantum yield of photobleaching and $\gamma_{exc}$ is the excitation rate of the UP (as shown in the previous Section, the resonant excitation of UP is the most harmful process for the molecules). We notice that the bleaching rate does not depend on $\gamma_{LP}$ as long as the LP relaxation is fast. Assuming $\gamma_{bl}$ does not depend on time the solution reads: $N(t) = N(0) e^{-\gamma_{bl} t}$. On the time scale of our experiments, the bleaching exhibits a quasi-linear behavior and thus can be approximated by $\frac{N(t)}{N(0)} \approx 1 - \gamma_{bl} t$, in agreement with experimental observations in Figs. 2-4. The bleaching rate $\gamma_{bl}$ can thus be directly assessed from experimental curves. In practice $\gamma_{bl}$ depends on the coupling strength, which in turn changes in time. Therefore, the slope of photochemical reaction should increase with time, as observed (see Fig. 2G). In our analysis we have, however, ignored this dependence for the sake of simplicity.

For the case of uncoupled J-aggregates in the absence of any nanoantenna, the bleaching rate is similarly given by: $\frac{dN_0}{dt} = -\gamma_{bl}^0 N_0$, where $N_0$ is the number of molecules in the uncoupled J-aggregate and $\gamma_{bl}^0 = \gamma_{exc}^0 \phi_{bl}^0$. Here, $\phi_{bl}^0 = \left(\frac{\gamma_{ISC}}{\gamma_0 + \gamma_{ISC}}\right)\left(\frac{k[^3O_2]}{k[^3O_2] + \gamma_T^0}\right)$ is the quantum yield of photobleaching and $\gamma_{exc}^0$ is the excitation rate for the uncoupled J-aggregates, correspondingly. The solution thus reads: $N_0(t) = N_0(0)e^{-\gamma_{bl}^0 t}$. On the time scale of the experiment this equation is exponential, which in turn implies that $\gamma_{bl}^0 \gg \gamma_{bl}$. This is in agreement with experimental observations in Figs. 2-4. The bleaching rate $\gamma_{bl}^0$ can thus be directly assessed from experimental exponential curves representing uncoupled J-aggregates.

To quantify the modification of the photobleaching dynamics, we introduce the Stabilization Factor (SF) as ratio between the bleaching rates of strongly coupled and uncoupled molecules:

$$SF = \frac{\gamma_{bl}^0}{\gamma_{bl}} = \frac{\gamma_{exc}^0 \phi_{bl}^0}{\gamma_{exc} \phi_{bl}} \quad \text{Eq. (2)}$$

Since the triplet state is far away from any other states, we assume that the term $\left(\frac{k[^3O_2]}{k[^3O_2] + \gamma_T}\right)$ does not significantly alter due to coupling and thus $\gamma_T^0 \approx \gamma_T$. We also assume that ISC rate is the same for uncoupled molecules and for the strong coupling case. Therefore, from Eqs. (1-2) it follows that in order to increase the molecular stability, that is increase the SF, $\gamma_{UP}$ and $\gamma_{D \to LP}$ must be maximized, while $\gamma_{UP \to D}$ and $\gamma_{exc}$ at the UP resonance minimized.

Importantly, it is possible to measure SF experimentally (see Fig. 5B,D). The data shows that SF varies between about ~200 for the highly red-detuned particles to about ~4 for highly blue-detuned particles. For the zero-detuned particles, SF varies between ~2 and ~18 depending on the Rabi splitting. The exact values depend on the parameters of the coupled system, which we evaluate further.

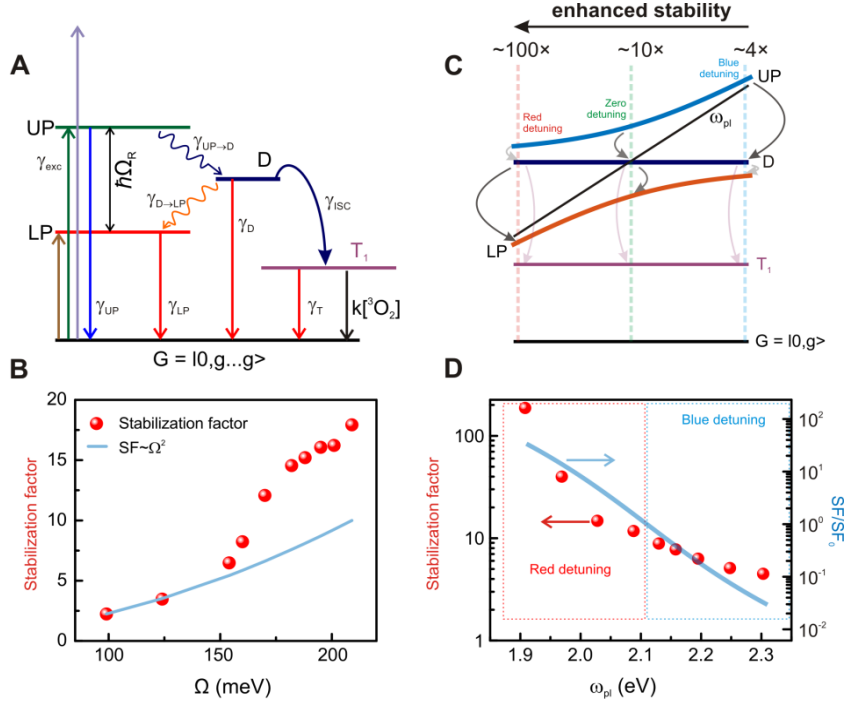

**Fig. 5. Scheme of photobleaching mechanism.** (A) Schematic representation of a strongly coupled system and transitions between various states. (B) Stabilization factor as a function of Rabi splitting energies. Symbols represent experimental results and are normalized by the bleaching rate of uncoupled J-aggregates. The solid blue line represents a quadratic dependence on Rabi splitting. (C) Schematic representation of photobleaching as a function of plasmon-exciton detuning. The visibility/transparency of arrows represents high/low probability of the corresponding transition, demonstrating that highly red-detuned particles are much more stable than highly blue-detuned ones. (D) Stabilization factor as a function of plasmon resonance frequency. Symbols represent experimental results and are normalized by the bleaching rate of uncoupled J-aggregates. The solid blue line represents the relative stabilization factor (normalized by the stabilization factor at zero detuning) calculated using Eq. 4 at $\Omega_R$=200 meV. Note the logarithmic scale.

To explain our experimental observations, we now turn to the relaxation dynamics and estimate the quantum yield of photobleaching in strong and weak coupling cases - $\phi_{bl}$ and $\phi_{bl}^0$. Dynamics of strongly coupled organic microcavities has been extensively studied theoretically by Litinskaya and Agranovich *et al.* (*46, 47*) using Fermi's golden rule approach, by Michetti *et al.* (*48, 49*) using numerical methods and more recently by Herrera *et al.* (*25, 50*) using the Holstein-Tavis-Cummings approach. From these works it follows that the direct contribution of upper and lower polaritons to photobleaching is likely negligible since their lifetime is extremely short (~10 fs), a few orders of magnitude shorter than typical rate of intersystem crossing $\gamma_{ISC}$ of organic molecules, which lies in the nanosecond to millisecond range depending on the strength of spin-orbit coupling (*44, 45*). The upper polariton, however, can rapidly decay (~50 fs) into incoherent molecular states via a phonon mediated process (*46*). There the population can reside for a much longer time (~10 ps) and thus the impact of these states on the photo-oxidation must be crucial. The transition rate from UP to D is given by (*46*):

$$\gamma_{UP \to D} \sim g^2 \omega_{vib}^2 \frac{(\Omega_R/2)^2}{(\omega_{UP}-\omega_0)^2+(\Omega_R/2)^2} e^{-(\omega_{UP}-(\omega_0+\omega_{vib}))^2/\gamma_0^2} \qquad \text{(Eq. 3)},$$

where $g^2$ is a dimensionless Huang-Rhys factor, which characterizes the strength of vibronic coupling, $\omega_{vib}$ is the phonon energy, $\omega_0$ is exciton resonance frequency and $\gamma_0$ is the total broadening of the excitonic resonance. The transition probability is maximized when $\omega_{UP} = \omega_0 + \omega_{vib}$ and thus depends on the density of available phononic modes in J-aggregates. Raman spectroscopy shows that the frequency of molecular vibrations spans over a broad range from 30 to ~430 meV (*54, 55*). At zero detuning, Eq. 1 reduces to $\gamma_{UP \to D} \sim g^2 \omega_{vib}^2 e^{-(\Omega_R/2-\omega_{vib})^2/\gamma_0^2}$, which shows that the optimum is achieved when $\frac{\Omega_R}{2} = \omega_{vib}$. In practice this implies that there will always be a phonon matching the corresponding transition, even for rather high Rabi splittings. The rate of transition is proportional to $g^2 \omega_{vib}^2$ and thus it is increased for high frequency phonons provided the vibronic coupling is not negligible.

The subsequent relaxation of incoherent states is also determined by the degree of strong coupling. In particular, the non-radiative transition from relatively long-lived incoherent states into a short-lived LP via emission of an intramolecular phonon is accelerated due to an increased Rabi splitting. In this case, the energy is transferred from the excited incoherent state into the lower polariton state with a single quantum of vibrational energy localized at the same molecule as the original electronic excitation. The matrix element of the corresponding exciton-phonon Hamiltonian is proportional to $\sim g\omega_{vib}$. Thus, the probability for such a transition is given by (*47*): $\gamma_{D \to LP} \sim <|a_s|^2> g^2 \omega_{vib}^2$, where $<|a_s|^2>$ is the contribution of the *s*th molecule into the LP state (averaged over all molecules). Therefore, photobleaching can be suppressed by depletion of the incoherent states population into the lower polariton where fast relaxation occurs. This depletion is especially accelerated by high-frequency phonons (see Fig. 5C).

With this at hand, we can now evaluate how the relaxation rates depend on detuning, Rabi splitting and excitation wavelength. Recalling Eqs. (1-2), $\gamma_{UP}$ and $\gamma_{D \to LP}$ must be maximized, while $\gamma_{UP \to D}$ and $\gamma_{exc}$ minimized. This is in line with our experimental observations. In particular, highly red-detuned particles are very stable due to efficient depopulation of the D states and relatively small $\gamma_{exc}$ due to a reduced plasmonic component of the UP (Fig. 5D). The highly blue-detuned particles are, on the contrary, very unstable because inefficient depopulation of D, efficient $\gamma_{UP \to D}$ and relatively high $\gamma_{exc}$ due to increased plasmonic component of the UP (Fig. 5D). Furthermore, the expression for SF can be simplified to yield $SF \propto \frac{\gamma_{UP}}{\gamma_{exc}} \frac{\omega_{vib(D \to LP)}^2}{\omega_{vib(UP \to D)}^2}$, assuming $\gamma_{UP} \gg \gamma_{UP \to D}$, $\gamma_{D \to LP} \gg \gamma_{ISC}$, and equality of Huang-Rhys factors for the involved phonons. Further, we note that the ratio $\frac{\gamma_{UP}}{\gamma_{exc}}$ is rather insensitive to detuning as $\gamma_{UP}$ and $\gamma_{exc}$ work against each other. In this simple case, stabilization is determined by the ratio of the phonon energies responsible for upper polariton to exciton reservoir and exciton reservoir to lower polariton relaxation rates. Since the phonon energy must be equal to the difference between $\omega_{UP}$ and $\omega_0$, and correspondingly $\omega_0$ and $\omega_{LP}$, then:

$$SF \propto \frac{(\omega_0 - \omega_{LP})^2}{(\omega_{UP} - \omega_0)^2} \qquad \text{(Eq. 4),}$$

This is shown in Fig. 5D as a blue solid line for $\Omega_R$=200 meV. Considering extreme simplifications made and the absence of any free parameters, the agreement with experiment is remarkable.

In case of zero-detuning, the phonon energies for the $UP \to D$ and $D \to LP$ transitions are equal. Hence, the relative stability is determined by the interplay of the UP excitation and decay rates: $SF \propto \frac{\gamma_{UP}}{\gamma_{exc}}$. For higher Rabi splitting this ratio increases, leading to formation of more stable hybrids. This is because of the coherent nature of polaritonic states, which implies that the relaxation rate of the UP depends on the number of involved excitons as $\gamma_{UP} = \frac{\gamma_{pl} + N\gamma_0}{2}$, while at the same time the excitation is shared between $N$ sites resulting in $\gamma_{exc} \propto \frac{\sigma_{ext}}{N}$ to the first approximation (see section S4). This in turn implies that for the case of zero-detuning $SF \propto \Omega^2$. In the experiment we observe even faster growth of the SF with the Rabi splitting (see Fig. 5B). This may happen because the white light excitation used here can also populate the exciton reservoir non-resonantly. The subsequent decay into the lower polariton is more efficient for higher Rabi splitting, making particles with large Rabi splitting more stable.

**DISCUSSION**

To conclude, we have demonstrated that strong coupling of organic chromophores to metallic nanostructures can significantly stabilize the former against photobleaching. This is achieved due to efficient suppression of triplet population by modified relaxation pathways in the strong coupling regime. In particular, involvement of coherent polaritonic states speeds up the relaxation, thus efficiently competing with the inter-system crossing and thereby blocking the photobleaching.

Stabilization against photobleaching in presence of strongly amplified electromagnetic fields may come as a surprise, as one could expect speed up, not slow down, of the reactions. Indeed, plasmonic nanoparticles and their arrays have been long known to boost photochemical reactions (*56-59*). In this "surface-enhanced photochemistry" regime, the excitation rate of the molecule is increased by the interaction with the antenna $\gamma_{exc} > \gamma_{exc}^0$, while the quantum yield of the corresponding photochemical process remains weakly perturbed $\phi \approx \phi^0$. In our experiments, even though enhanced fields are present, $\gamma_{exc} > \gamma_{exc}^0$, the quantum yield of photobleaching is significantly reduced, $\phi_{bl} \ll \phi_{bl}^0$, due to involvement of collective UP and LP states. Therefore, the coupled system can be significantly stabilized.

In addition, when molecules reside very close to metal surface, their electronic orbitals may mix with the orbitals of the metal resulting in charge transfer processes and renormalization of HOMO-LUMO gaps (*60*). Such processes can play an important role in surface chemistry. In addition to that, hot electrons (and holes) can affect the molecular stability and in particular damage adsorbed molecules by populating their LUMO states (*61-63*). While it is hard to completely rule out the hot-electron scenario, we notice that hot electrons should additionally destabilize the molecules. In our experiments, the coupled system is on the contrary significantly more stable than uncoupled J-aggregates. We thus argue that hot-electrons effects are likely minor. This may be explained by protective ligand layer that is present between Ag nanoprisms and J-aggregates and that may hinder direct charge transfer (see Methods). In addition, because of the relatively large size of Ag nanoprisms used in this study, hot electrons are likely to thermalize within the metallic particle before reaching the surface.

In our study, the incoherent states play a major role, as they are the only long-lived intermediates capable of populating the triplet states. We note, however, that in a situation where no long-lived intermediates exist, the hybrid system in the strong coupling regime could become completely immune to photobleaching. This hypothesis could be experimentally verified in the single molecule strong coupling case, where no incoherent states exist (*64*). We also note that in our study the triplet state lies far below the LP state and thus its dynamics is not directly affected by the coupling. However, the coupling potentially can be made so strong that the triplet states and LP can be co-aligned. This will enhance $\gamma_T$ via activation of back ISC process from triplet to the LP. While this was not the case here, it can in principle be utilized to further increase the stability of such samples. Similar ideas have been previously verified both experimentally and theoretically in the weak coupling regime (*38-40*).

More generally, our findings indicate that in the strong coupling regime the kinetics of reactions can be dramatically modified in comparison to the corresponding uncoupled case. This in turn implies that for an arbitrary chemical reaction at equilibrium, where only one of the involved substances is strongly coupled to the vacuum field of the cavity, the equilibrium will be affected in favor of stabilization of the strongly coupled product (*23-25*).

In conclusion, our findings confirm that strong light-matter interaction plays an important role in determining the material properties of coupled systems and may have far reaching implications for the development of nanophotonic devices incorporating organic molecules. We note that similar stabilization mechanisms can be important for other classes of molecules, such as perovskites and/or improved organic electronic devices where triplet population is undesirable. We also anticipate that stabilization by strong coupling can improve the performance of various organic opto-electronic devices, such as solar cells, organic light emitting diodes, and others.

**MATERIALS AND METHODS**

*Materials:* Silver nitrate (AgNO$_3$, 99.9999%), trisodium citrate dihydrate, potassium bromide (KBr), sodium borohydride (NaBH$_4$, 99%), and Bis(p-sulfonatophenyl)phenylphosphine dehydrate dipotassium salt (BSPP) were purchased from Sigma-Aldrich at the highest purity grade available. 5,6-Dichloro-2-[[5,6-dichloro-1-ethyl-3-(4-sulfobutyl)-benzimidazol-2-ylidene]-propenyl]-1-ethyl-3-(4-sulfobutyl)-benzimidazolium hydroxide, inner salt, sodium salt (TDBC, purchased from FEW chemicals) was used without further purification. TEM grids were purchased from Ted Pella, Inc. All glassware and stir bars were thoroughly pre-cleaned and dried prior to use. Ultra-pure distilled water (Millipore, 18 MΩ cm) was used in all preparations.

*Synthesis of silver nanoprisms:* Silver nanoprisms (AgNPs) were synthesized according to the literature (*65*). In brief, a 3-5 nm silver seed nanoparticles solution was prepared by adding 1 mL of 30 mM trisodium citrate and 0.5 mL of 20 mM silver nitrate (AgNO$_3$) solutions to 95 mL of an ice-cold ultra-pure distilled water. The solution was kept bubbling with N$_2$ under vigorous stirring in an ice-bath for an additional 60 min in the dark. Then, 1 mL of ice-cold 50 mM NaBH$_4$ was added into the growth solution, at which point the color of the solution turns pale-yellow. Subsequently, 100 μL of 50 mM NaBH$_4$ was added to the solution, and repeated 3 more times with 2 min break in between. The ice-cold and freshly prepared mixture of 1 mL of 5 mM BSPP and 1 mL of 50 mM NaBH$_4$ was added dropwise into the growth solution for seed nanoparticles. The solution was kept for 5 h in an ice bath under gentle stirring and completed with aging overnight in an ice bath. For the photo-induced growth of AgNPs, 10 mL of the aged seed solution with pH of 9.5 was irradiated with a 532 nm continuous-wave (CW) laser. The reaction was allowed to proceed for 24 h, at which point the AgNPs were washed two times by centrifugation at 3000 rcf for 5 min, discarding the supernatant, and re-dispersed in aqueous solution containing 0.3 mM of trisodium citrate.

*Synthesis of the hybrid nanostructures:* For the hybrid nanostructures, consist of self-assembled J-aggregates of TDBC dye molecules on the individual AgNPs, we followed a previously reported method with slight change (*10, 66*). The freshly synthesized AgNPs (1mL) were capped by mixing with 1 mL of the aqueous solution containing 0.1 mM TDBC dye molecules and 1 mM of KBr under gentle stirring for 15 min. The resulting colloidal solution was washed once by centrifugation (3000 rcf for 5 min), re-dispersed in distilled water, and stored at 4ºC until use.

*Optical spectroscopy:* The sample was prepared by drop casting the 3 μL of hybrid nanoparticles solution on a polylysine-functionalized transmission electron microscope (TEM, 200 mesh) grid (90×90 μm$^2$). After 2 min, the solution was removed with dust-free tissue, and dried with gentle nitrogen flow. Using a transparent and conductive TEM grid allows for spectroscopic and morphological correlation. Bare J-aggregates sample was prepared from the solution containing 0.1 mM TDBC dye molecules and 1 mM of KBr with same procedure as for the hybrid nanoparticles on TEM grid. DF scattering spectra were recorded using an inverted microscope (Nikon TE-2000E equipped with an oil-immersion 100×/NA 0.5-1.3 objective) with a hyper-spectral imaging technique based on a liquid crystal tunable filter (*13*). The morphology characterization of hybrid nanoparticles was analyzed using a scanning electron microscope (Ultra 55 FEG SEM).

*Photo-bleaching experiment:* Photo-bleaching experiments under flat-broadband illumination were conducted with an irradiance of ~15 W/cm$^{-2}$ at the sample using a laser-driven white light source (LDLS, EQ-99FC, high-brightness, flat-broadband spectrum) with a visible filter (350-715 nm, Thorlabs) in normal atmospheric condition at room-temperature. DF scattering spectra were recorded before and after iteratively 2 minutes of continuous photo-bleaching. The relative change of the DF scattering spectra was used to estimate photo-bleaching rate of the samples over 10 minutes using a linear fitting.

We note that photobleaching using narrow excitation in Fig. 4 was considerably slower than broad range source in Figs. 2-3. This is because the integrated power of the broad range source was significantly higher. In this experiment, three different band pass filters with central wavelengths of 532, 568, and 640 nm were used for the photo-bleaching experiment with different excitation resonances of 532, 568, and 640 nm, respectively. The photo-bleaching rates for the different excitation resonances were normalized by the total irradiance to make a fair comparison.

To evaluate the contribution of molecular oxygen to the photobleaching in organic dye molecules, we performed a control experiment in the vacuum chamber (see fig. S3 in the section S3). The photobleaching experiment was performed using a 532 nm continuous wave laser excitation (~200 W/cm$^{-2}$) in an optical cryostat vacuum chamber and a long working distance objective (Nikon, 20× NA=0.45). Dark-field (DF) scattering spectra of pristine J-aggregates on a silicon substrate, inside and outside of an optical cryostat vacuum chamber, are recorded before and after continuous photobleaching. While J-aggregates in the presence of oxygen is rapidly photobleached, they are highly stable and show nearly unaffected optical response inside the vacuum chamber. We thus conclude that atmospheric oxygen plays a crucial role in our photobleaching experiments.

*Numerical Calculations:* Finite-difference time-domain (FDTD) calculations were used to compute scattering and absorption spectra of silver nanoprisms coupled to J-aggregates. The dimensions of simulated nanoprisms were tuned using SEM images as guidance to obtain a plasmon resonance overlapping with the J-aggregate absorption line as well as red- and blue-shifted. The nanoprisms for zero-detuning were 75 nm in length (side) with a corner rounding of 9 nm and edge rounding of 3 nm. The thickness was in this case 19 nm. Red-shifted nanoparticles had their thickness reduced to 13 nm, what is supported by SEM images indicating similar size (to zero-detuned particles) with a different contrast due to being thinner. In the case of blue-shifted nanoprisms the thickness was increased to 29 nm and the side length reduced to 70 nm. These geometrical parameters with permittivity of silver taken from Palik (*67*) yielded, together with a Lorentzian permittivity for J-aggregate, scattering spectra consistent with experimental measurements. The J-aggregate was modelled in every case by a Lorentzian permittivity function with a background refractive index of 1.45, absorption line at 590 nm, reduced oscillator strength of 0.07, and linewidth of 50 meV. The thickness in each case was uniform and equal to 2 nm, what is consistent with a single monolayer of J-aggregate. The particles were placed on a glass substrate with n=1.45.

The particles were excited by a plane wave in the total-field/scattered-field configuration, what is appropriate for such thin particles whose out-of-plane resonance excited in a dark-field configuration is significantly detuned from the transverse plasmon. Total scattering and absorption cross sections were computed from the Poynting vector, while absorption in individual parts of the structure – the silver and the J-aggregate separately – was based on the imaginary part of the permittivity and the electric field intensity as a particular point inside either of the two absorptive media. To assure convergence a fine mesh of 0.5 nm was used around the nanoparticle.

**SUPPLEMENTARY MATERIALS**

# Suppression of photo-oxidation of organic chromophores by strong coupling to plasmonic nanoantennas

Battulga Munkhbat[1], Martin Wersäll[1], Denis G. Baranov[1], Tomasz J. Antosiewicz[1,2] and Timur Shegai[1*]

1. Department of Physics, Chalmers University of Technology, 412 96, Göteborg, Sweden

2. Centre of New Technologies, University of Warsaw, Banacha 2c, 02-097 Warsaw, Poland

e-mail: timurs@chalmers.se

*section S1. Evolution of the dark-field scattering spectra of strongly coupled hybrid systems with different initial Rabi splitting energies.*

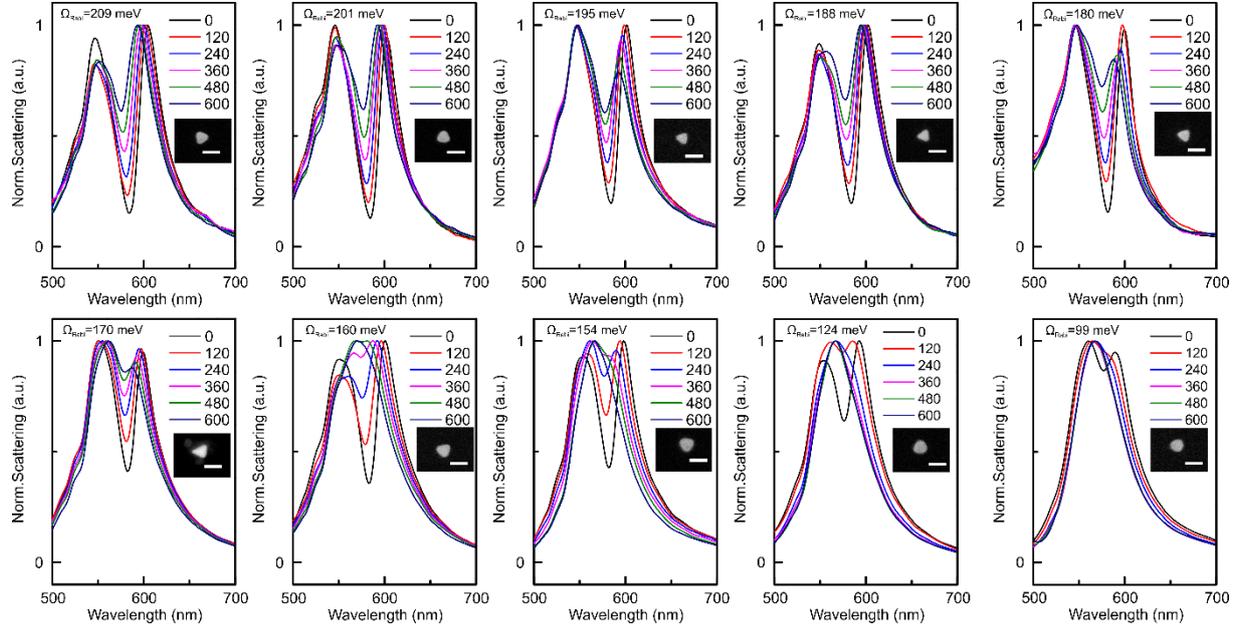

**fig S1.** Evolution of the dark-field scattering spectra of strongly coupled hybrid systems with different initial Rabi splitting energies. Insets show the corresponding SEM images of the plasmonic silver nanoprisms. The scale bar is 100 nm.

*section S2. Evolution of the dark-field scattering spectra of strongly coupled hybrid systems with different detuning of plasmon resonance with respect to the exciton resonance.*

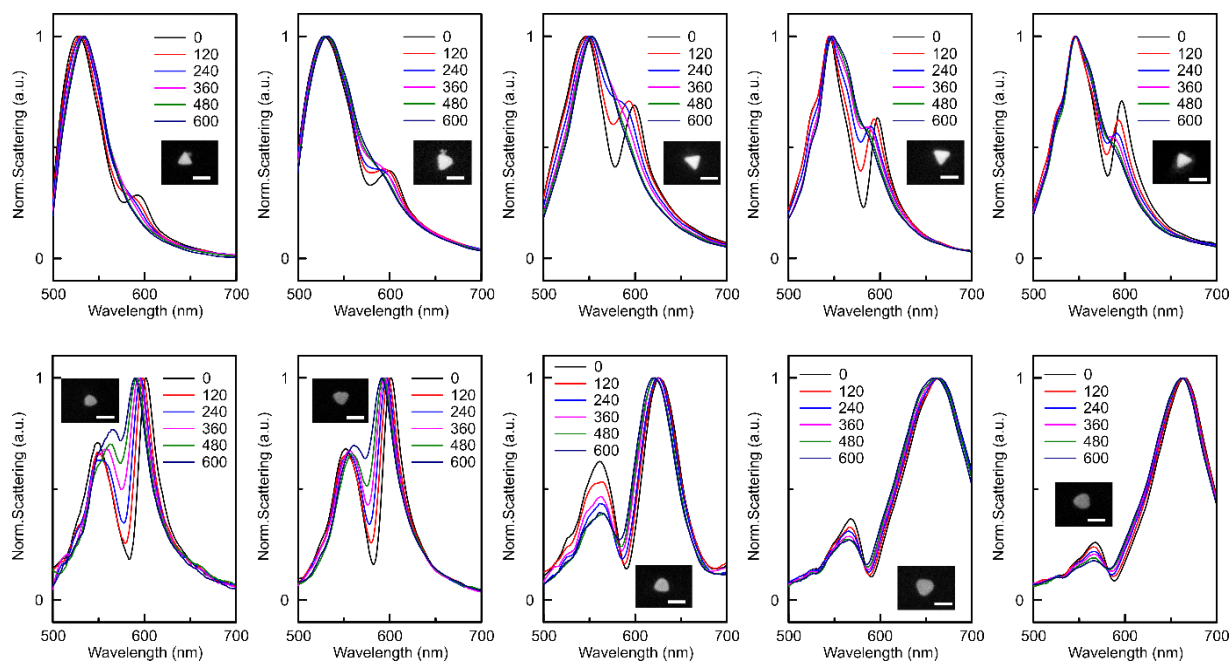

**fig S2.** Evolution of the dark-field scattering spectra of strongly coupled hybrid systems with different detuning of plasmon resonance with respect to the exciton resonance. Insets show the corresponding SEM images of the plasmonic silver nanoprisms. The scale bar is 100 nm.

*section S3. Contribution of Molecular Oxygen for Photobleaching.*

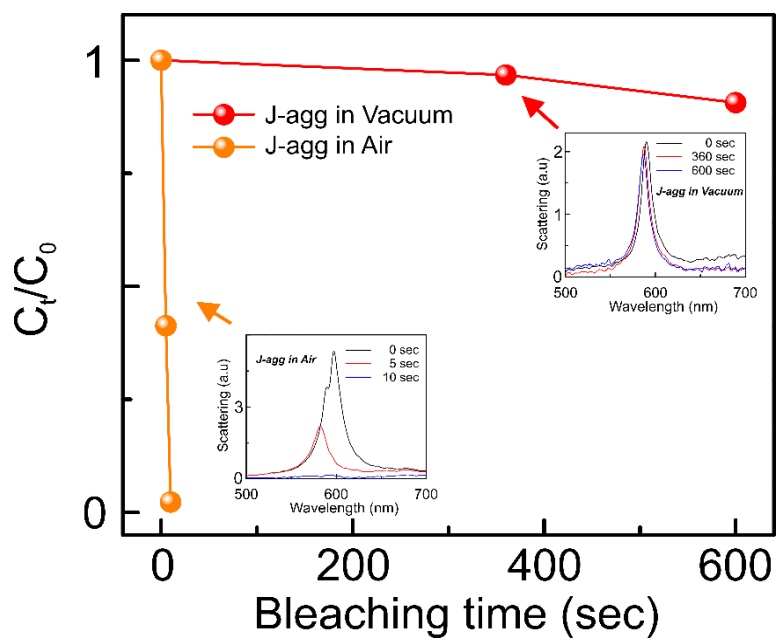

**fig S3.** Contribution of Molecular Oxygen for Photobleaching. Photobleaching kinetics for J-aggregates in air (orange) and in vacuum (red) under 532 nm laser excitation. The insets show the corresponding DF spectra of J-aggregates in air and in vacuum.

To evaluate the contribution of molecular oxygen to the photobleaching in organic dye molecules, we performed a control experiment, as is shown in fig. S3. The photobleaching experiment is performed with a 532 nm laser excitation (~200 W/cm$^{-2}$) using an optical cryostat vacuum chamber and a long working distance objective (Nikon, 20× NA=0.45) in an upright microscope. DF scattering spectra of bare J aggregates on a silicon substrate, inside and outside of an optical cryostat vacuum chamber, are recorded before and after continuous photobleaching. While J-aggregates in the presence of oxygen is rapidly bleached within a few seconds under the laser illumination, the one inside cryostat vacuum chamber shows almost same intensity as before photo-bleaching. (see fig. S3). Obviously, it is indeed that molecular oxygen plays a major role in the photobleaching.

*section S4. Coupled system rate equations:*

The rate equations given below describe the temporal dynamics of the coupled system:

$$\frac{dn_G}{dt} = -\gamma_{exc} n_G + \gamma_{UP} n_{UP} + \gamma_D n_D + \gamma_{LP} n_{LP} + \gamma_T n_T \qquad \text{Eq. (S1)}$$

$$\frac{dn_{UP}}{dt} = \gamma_{exc} n_G - (\gamma_{UP \to D} + \gamma_{UP}) n_{UP} \approx 0 \qquad \text{Eq. (S2)}$$

$$\frac{dn_D}{dt} = \gamma_{UP \to D} n_{UP} - (\gamma_{D \to LP} + \gamma_D + \gamma_{ISC}) n_D \approx 0 \qquad \text{Eq. (S3)}$$

$$\frac{dn_{LP}}{dt} = \gamma_{D \to LP} n_D - \gamma_{LP} n_{LP} \approx 0 \qquad \text{Eq. (S4)}$$

$$\frac{dn_T}{dt} = \gamma_{ISC} n_D - (\gamma_T + k[^3O_2]) n_T \approx 0 \qquad \text{Eq. (S5)}$$

$$\frac{dn_P}{dt} = k[^3O_2] n_T \qquad \text{Eq. (S6)}$$

here $n_i$ denotes the *population* of the $i^{th}$ state ($i$=G, UP, D, LP, T and P – the photoproduct) and $\gamma_i$ denotes the corresponding transition rate. The condition of quasi-stationarity implies that changes in population of all intermediates occur very slowly, for which reason Eqs. (S2-S5) are all set to zero. This is justified when the excitation rate is much slower than any other relevant rate (weak excitation condition). In addition to that, the total population obeys the normalization condition: $\sum_i n_i = 1$, from which, together with Eqs. (S1-S6), it follows that: $\frac{dn_G}{dt} = -\gamma_{bl} n_G$. Here $\gamma_{bl} = \gamma_{exc} \phi_{bl}$ is the photobleaching rate and $\phi_{bl}$ is the photobleaching quantum yield. Thus the population of the ground state is slowly transferred to the population of the photoproduct as the bleaching proceeds. The population of the ground state can be determined in terms of number of molecules participating in the coupling process as $n_G(t) = \frac{N(t)}{N(0)}$. This parameter can be directly deduced from the Rabi splitting, which in turn implies: $\frac{dN}{dt} = -\gamma_{exc} \phi_{bl} N$. Here the excitation rate, $\gamma_{exc}$, has a physical meaning of excitation probability per unit time *per molecule*. Having that in mind, we proceed to the derivation of $\gamma_{exc}$.

We first obtain a general expression for the excitation rate of a single-mode optical cavity illuminated by a plane wave of intensity $I_0$. Let $\omega_0$ be the mode frequency, $\gamma_{rad}$ the elastic (radiative) decay rate of the mode and $\gamma_{nr}$ the rate of all inelastic (including non-radiative) decay processes. According to the temporal coupled modes theory, the power scattered and absorbed by the cavity mode reads $P_{sca} = \gamma_{rad} n_{st} \hbar \omega_0$ and $P_{abs} = \gamma_{nr} n_{st} \hbar \omega_0$, correspondingly. Here, $n_{st}$ is the stationary population of the excited state of the cavity. Adding up these two expressions, we obtain:

$$P_{ext} = P_{sca} + P_{abs} = (\gamma_{rad} + \gamma_{nr}) n_{st} \hbar \omega_0.$$

On the other hand, $P_{ext} = \sigma_{ext} I_0$ with $\sigma_{ext}$ being extinction cross-section of the cavity. Combining this with the rate equation for the mode population $\dot{n} \approx \gamma_{exc} - (\gamma_{rad} + \gamma_{nr}) n$, we obtain the final expression for the excitation rate: $\gamma_{exc} = \sigma_{ext} I_0 / \hbar \omega_0$. From here, we recall that we wish to calculate the excitation rate per molecule. Thus, the original expression $\gamma_{exc} = \sigma_{ext} I_0 / \hbar \omega_0$ has to be normalized by the number of participating molecules, which results in $\gamma_{exc} = \frac{\sigma_{ext} I_0}{N \hbar \omega_0}$.